\documentclass[aps,prd,longbibliography,nofootinbib,amsthm,amsmath,amssymb,amsfonts,notitlepage]{revtex4-1}
\usepackage[utf8]{inputenc}
\usepackage[T1]{fontenc}
\usepackage[english]{babel}
\usepackage{graphicx}
\usepackage{float}
\usepackage{xcolor}
\usepackage{booktabs}

\renewcommand{\Im}{\mathrm{Im }}
\newcommand{\doublet}[2]{ \left( \begin{array}{c}#1 \\ #2 \end{array}\right) }

\newcommand{\triplet}[3]{ \left(\! \begin{array}{c}#1 \\ #2 \\ #3 \end{array}\!\right) }

\newcommand{\lr}[1]{ \langle #1 \rangle}

\newcommand{\mmmatrix}[9]{ \left(\!\! \begin{array}{ccc}#1 & #2 & #3\\ #4 & #5 & #6\\ #7 & #8 & #9\\ \end{array}\!\!\right) }

\newcommand{\hsm}{h_{\mbox{\tiny SM}}}
\newcommand{\hnew}{h_{\mbox{\footnotesize  146}}}

\newcommand{\hmutau}{\mathcal{B}_{\mu\tau}^h}
\newcommand{\mueg}{\mathcal{B}_{e\gamma}^\mu}
\newcommand{\mueee}{\mathcal{B}_{3e}^\mu}
\newcommand{\RomanNum}[1]{\expandafter\@slowromancap\romannumeral #1@}

\newcommand{\msm}{m_{\mbox{\tiny SM}}^2}

\newcommand{\GeV}{\,\mbox{GeV}}
\newcommand{\CP}{\textit{CP} }
\newcommand{\Vpmns}{V_{\text{{\tiny PMNS}}}}
\newcommand{\Br}{\mathrm{Br}}

\newcommand{\lsim}{\mathrel{\rlap{\lower4pt\hbox{\hskip1pt$\sim$}}
		\raise1pt\hbox{$<$}}}         
\newcommand{\gsim}{\mathrel{\rlap{\lower4pt\hbox{\hskip1pt$\sim$}}
		\raise1pt\hbox{$>$}}}         
	
\definecolor{darkred}{rgb}{0.8,0.1,0.1}
\definecolor{paleyellow}{rgb}{1.0,1.0,0.7}

\begin{document}
	
\title{Lepton flavor violating signals driven by CP symmetry of order 4}

\author{Bei Liu}
\email{liub98@mail2.sysu.edu.cn}
\author{Igor P. Ivanov}
\email{ivanov@mail.sysu.edu.cn}
\affiliation{School of Physics and Astronomy, Sun Yat-sen University, 519082 Zhuhai, China}

\begin{abstract}
CP4 3HDM is a curious version of the three-Higgs-doublet model built upon a CP symmetry of order 4 (dubbed CP4).
When extended to fermions, CP4 leads to unusually tight correlations between the scalar and Yukawa sectors 
and induces tree-level flavor changing neutral couplings. Still, viable scenarios exist, in which quark flavor changing signals
remain within experimental limits. In this work, we extend CP4 to the lepton sector and 
investigate whether the lepton-Higgs couplings and lepton flavor violating (LFV) signals can also be kept under control. 
We consider two classes of LFV processes: 
tree-level leptonic decays of the 125 GeV Higgs boson and the muon decays $\mu\rightarrow e\gamma$ and $\mu \to 3e$. 
For each CP4-invariant lepton Yukawa scenario, we perform a focused Yukawa sector scan that uses physical lepton properties
as input and suppresses LFV effects. 
We identify a promising CP4 3HDM scenario compatible with the present-day experimental
constraints, show that it can accommodate the recent CMS hint of a 146 GeV scalar decaying to $e\mu$, 
and argue that this interpretation can be tested at future colliders through the search of comparably strong $ee$ and $\mu\mu$
signals.
\end{abstract}

\maketitle

\section{Introduction}
Among many possible directions for building models beyond the Standard Model (BSM),
multi-Higgs-doublet models remain a very popular framework that can offer remarkably rich BSM phenomenology with a minimum of assumptions. This research direction began half a century ago with T.D.~Lee's construction of the two-Higgs-doublet model (2HDM) \cite{Lee:1973iz} and S.~Weinberg's suggestion of the three-Higgs-doublet model \cite{Weinberg:1976hu}. Both models were proposed as possible explanations for the origin of $CP$ violation. However, it was later recognized that such models offer interesting opportunities for flavor model building and can lead to characteristic collider signatures as well as peculiar cosmological consequences. As a result, the 2HDM and 3HDM literature now counts thousands of papers and continues to grow rapidly, see reviews \cite{Branco:2011iw,Ivanov:2017dad}.

In the last decade, the exploration of 3HDMs picked up pace as they offer numerous opportunities that cannot be realized within the 2HDM or singlet extensions of the scalar sector. In particular, 3HDMs can be built upon various global symmetries, either exact or broken \cite{Ivanov:2011ae,Ivanov:2012fp,Ivanov:2014doa,Darvishi:2019dbh,Kuncinas:2025uty}, which lead to characteristic signals worth exploring. When building such models, one usually tries to find a balance between the flexibility of the model and its predictive power. Models with a large symmetry content typically possess very few free parameters, are highly predictive, but conflict with experimental data. By relaxing symmetry assumptions, for example, by introducing soft breaking terms, one often succeeds in fitting experimental observations; however, with many additional free parameters, the model becomes less predictive and loses its theoretical appeal.

In \cite{Ivanov:2015mwl}, a peculiar version of the 3HDM was constructed, built on a single symmetry yet displaying a remarkably constrained scalar sector with unusual properties. The symmetry in question is a $CP$ transformation of order 4, CP4 for short. Although higher-order $CP$ transformations have been studied for decades \cite{Ecker:1987qp,Grimus:1995zi,Branco:1999fs}, this was the first example of a multi-Higgs model, in which the imposition of CP4 did not lead to any accidental symmetry, as it happened, for instance, in the 2HDM \cite{Maniatis:2007de,Ferreira:2009wh,Maniatis:2009vp}. Several phenomenological studies of the CP4 3HDM followed, focusing on the scalar sector \cite{Ferreira:2017tvy,Liu:2024aew}, the quark Yukawa sector \cite{Ferreira:2017tvy,Zhao:2023hws}, or the Higgs-top-quark couplings \cite{Ivanov:2021pnr,Zhao:2025rds}. The main message that emerged from these studies is the following. Out of eight possible CP4-invariant quark scenarios, only one survives after the $K$, $B$, $B_s$, and $D$-meson oscillations as well as top-quark couplings are taken into account \cite{Zhao:2025rds}. Admittedly, those works did not include all the experimental data available at present; thus, it remains an open question whether fully viable benchmark versions of the CP4 3HDM exist at all.

Our experience with this model shows that before embarking on a comprehensive scan of the vast parameter space and check numerous observables, we should focus on specific sectors and eliminate Yukawa scenarios that are already ruled out by a small set of signals and constraints. This step-by-step logic has already led to a much faster and more reliable scanning procedure in the scalar and the quark sectors.
However, the lepton sector of CP4 3HDM remains unexplored.

In this work, we explore for the first time whether extending CP4 to the lepton sector inevitably generates strong lepton-flavor-violating (LFV) signals that would be immediately ruled out by the present experimental constraints. It is well known that LFV processes are strongly suppressed in the SM and, therefore, are among the most sensitive probes of BSM physics, see review in \cite{Bernstein:2013hba, Lindner:2016bgg, Ardu:2022sbt, Perrevoort:2023jry}. Current experimental constraints on LFV are exceptionally tight. The MEG II experiment has set an upper limit of $\Br(\mu\rightarrow e\gamma < 1.5 \times 10^{-13})$ \cite{MEGII:2025gzr}, and future experiments, such as Mu2e \cite{COMET:2025sdw}, will further tighten these constraints. The Particle Data Group lists strict bounds on various LFV decays of the SM-like Higgs boson \cite{ParticleDataGroup:2024cfk}, and one can expect these upper bounds to go down as the ATLAS and CMS detectors accumulate more data.
Thus, it is necessary to check whether any of the CP4-invariant lepton sectors with non-trivially transforming leptons have a chance of avoiding conflict with the existing data and which LFV signatures can be detected in future experiments.

In this work, we address these questions by extending CP4 to the lepton sector and performing a systematic phenomenological analysis of selected LFV signals. We explore the parameter space using the special scanning procedures developed previously for the scalar \cite{Liu:2024aew} and Yukawa \cite{Zhao:2023hws} sectors of the model; these procedures use several physical observables as input, which strongly boosts the efficiency of the scan compared to the traditional approach of scanning the parameter space of the lagrangian.
For each of the three CP4-invariant Yukawa cases, we calculate the tree-level lepton decays of $\hsm$, both flavor-conserving and flavor-violating, and one-loop radiative decay $\mu\rightarrow e\gamma$, which can be mediated by both neutral and charged Higgs bosons. 

The outline of this paper is as follows. In section~\ref{section-LFV}, we review the scalar and lepton Yukawa sectors of the CP4 3HDM and list formulas used in numerical calculations. Section~\ref{section-numerical} presents numerical scans compared against experimental data. These results single out a specific CP4 3HDM lepton sector that passes all the constraints by a large margin. In section~\ref{section-146} we show that the recent CMS hint at a leptonically decaying 146~GeV scalar can be accommodated and tested within this model. We finish by summarizing our results and provide auxiliary information in the appendices.

Throughout the paper, the fermion generations are labeled as $i,j$, while the three Higgs doublets are denoted with the subscripts $a,b$. 
When dealing with the five neutral Higgs bosons, we label them with $k=1, \dots, 5$. With a slight abuse of notation, we use the same label $k$ for the two charged Higgs bosons, with $k = 1,2$. We believe that the choice is always clear and should not lead to confusion.

\section{LFV processes in the CP4 3HDM}\label{section-LFV}
\subsection{The scalar sector}

For the completeness, we remind the reader of the basics of non-standard \CP transformations and the CP4 symmetry.

In quantum field theory with several complex fields, the \CP transformations is not defined uniquely \cite{feinberg-weinberg,Lee:1966ik,Branco:1999fs,weinberg-vol1}. In a theory containing $N$ complex scalar fields $\phi_a, a=1,\dots,N$,
one can define a general \CP transformation (GCP) acting on the fields by $\phi_a \mapsto X_{ab} \phi_b^*$, with a unitary matrix $X$
(the spatial coordinate inversion is implicitly assumed). Applying such a \CP transformation twice, one generates a family transformation 
$\phi_a \mapsto (X X^*)_{ab}\phi_b$, which does not need to be the identity transformation.
It can happen that the identity is reached only after imposing the \CP transformation $2k$ times:
$(X X^*)^k = \mathbb{I}$; in this case, we say the the \CP transformation is of order $2k$. The usual \CP is of order 2, and the first non-standard example of a \CP transformation of order 4 (CP4 for short).

It is known that any CP4 transformation acting on three scalar fields can be represented, in a suitable basis, by the following transformation law \cite{Ecker:1987qp,Grimus:1995zi,Branco:1999fs}: $\phi_1 \mapsto \phi_1^*$, $\phi_2 \mapsto i \phi_3^*$,
$\phi_3 \mapsto -i\phi_2^*$, see Appendix~\ref{CP4-defs}:
The 3HDM scalar potential invariant under this transformation was constructed in \cite{Ivanov:2015mwl} as a sum of two parts $V = V_0+V_1$, 
where $V_0$ is invariant under all phase rotations of the doublets,
\begin{eqnarray}
	V_0 &=& - m_{11}^2 (\phi_1^\dagger \phi_1) - m_{22}^2 (\phi_2^\dagger \phi_2 + \phi_3^\dagger \phi_3) \nonumber
	+ \lambda_1 (\phi_1^\dagger \phi_1)^2 + \lambda_2 \left[(\phi_2^\dagger \phi_2)^2 + (\phi_3^\dagger \phi_3)^2\right]\\
	&&+ \lambda_{34} (\phi_1^\dagger \phi_1) (\phi_2^\dagger \phi_2 + \phi_3^\dagger \phi_3)
	- \lambda_4 \left[(\phi_1^\dagger \phi_1) (\phi_2^\dagger \phi_2) - (\phi_1^\dagger \phi_2)(\phi_2^\dagger \phi_1)
	+ (\phi_1^\dagger \phi_1) (\phi_3^\dagger \phi_3) - (\phi_1^\dagger \phi_3)(\phi_3^\dagger \phi_1)\right]\nonumber\\
	&&+
	\lambda'_{34} (\phi_2^\dagger \phi_2) (\phi_3^\dagger \phi_3)
	- \lambda'_4 \left[(\phi_2^\dagger \phi_2) (\phi_3^\dagger \phi_3) - (\phi_2^\dagger \phi_3)(\phi_3^\dagger \phi_2)\right]\,,
	\label{V0}
\end{eqnarray}
with all parameters real, while $V_1$ is given by
\begin{equation}
	V_1 = \lambda_5 (\phi_3^\dagger\phi_1)(\phi_2^\dagger\phi_1) +
	\lambda_8(\phi_2^\dagger \phi_3)^2 + \lambda_9(\phi_2^\dagger\phi_3)(\phi_2^\dagger\phi_2-\phi_3^\dagger\phi_3) + h.c.
	\label{V1a}
\end{equation}
with a real $\lambda_5$ and complex $\lambda_8$, $\lambda_9$. 
The notation used here follows the choice made in the most recent papers \cite{Liu:2024aew,Zhao:2025rds}.
We stress that the CP4 symmetry is exact; no soft breaking terms are introduced.

We assume that the parameters are such that minimum of potential is neutral and breaks CP4 symmetry.
The latter requirement is needed because an unbroken global symmetry acting on scalars that participate in 
fermion mass generations will unavoidably lead to pathological quark sector \cite{GonzalezFelipe:2014mcf}.
The three initial Higgs doublets acquire vacuum expectation values (vevs) that may be complex. 
However, as shown in \cite{Ferreira:2017tvy}, after a suitable basis transformation that preserves the form of the potential, 
these vevs can be made real:
\begin{equation}
	\lr{\phi_a^0} = 
	\frac{v}{\sqrt{2}} (c_\beta,\, s_\beta c_\psi,\, s_\beta s_\psi) \,.
	\label{vevs1}
\end{equation}
with $v=246 \GeV $ and shorthand notation $c_x \equiv \cos(x), s_x \equiv \sin(x)$. Thus, the position of the minimum is parameterized by two angles $\beta$ and $\psi$.

Description of the scalar sector becomes more transparent 
if we pass from the original basis with the doublets $\phi_a$ to a Higgs basis with $\Phi_a$, where only one doublet gets a non-zero vev:
$\lr{\Phi^0_1} = v/\sqrt{2}$, $\lr{\Phi_2} = \lr{\Phi_3} = 0$. Following the notation of \cite{Zhao:2023hws,Liu:2024aew}, we use the following transformation:
\begin{equation}
	\triplet{\Phi_1}{\Phi_2}{\Phi_3} =
	\mmmatrix{c_\beta}{s_\beta c_\psi}{s_\beta s_\psi}{-s_\beta}{c_\beta c_\psi}{c_\beta s_\psi}{0}{-s_\psi}{c_\psi}
	\triplet{\phi_1}{\phi_2}{\phi_3}\,. 
\end{equation} 
In this Higgs basis, we expand the doublets as
\begin{equation}
	\Phi_1 = {1\over\sqrt{2}}\doublet{\sqrt{2}G_1^+}{v + \rho_1 + i G^0},\
	\Phi_2 = {1\over\sqrt{2}}\doublet{\sqrt{2}w_2^+}{\rho_2 + i \eta_2},\
	\Phi_3 = {1\over\sqrt{2}}\doublet{\sqrt{2}w_3^+}{\rho_3 + i \eta_3}.\label{expansion-Higgs-basis}
\end{equation}
with $G^0, G^\pm$ denoting the would-be Goldstone bosons. 
The five neutral and two charged fields can be organized as
\begin{equation}
	\Psi_k = (\rho_1, \rho_2, \rho_3, \eta_3, \eta_2)\,, \quad
	\Psi_k^\pm = (w_2^\pm, w_3^\pm)\,.
\end{equation}
We remind the reader that $k=1, \dots, 5$ when labeling the neutral Higgs bosons and $k=1,2$ when denoting the charged Higgs bosons.
The order of last two neutral components in $\Psi_k$ (that is, $\eta_3, \eta_2$, not $\eta_2, \eta_3$) is motivated by the tridiagonal form of the scalar mass matrix derived in \cite{Liu:2024aew}. These fields are not physical scalars, and both mass matrices (in the neutral and charged sectors) must be further diagonalized by appropriate rotations \cite{Liu:2024aew,Zhao:2025rds}.

In the exact scalar alignment, $\rho_1$ would be identical to the SM like Higgs boson $\rho_1 = \hsm$.
However, the model does not force us to stick to the exact alignment. Following \cite{Liu:2024aew,Zhao:2025rds},
we allow for a small amount of misalignment.
A convenient way to introduce it is to parametrize $\hsm$ as
\begin{equation}
	\hsm = \rho_1 c_\epsilon + s_\epsilon c_\alpha (c_{\gamma_1} \rho_2 + s_{\gamma_1} \rho_3) +  
	s_\epsilon s_\alpha (c_{\gamma_2} \eta_3 + s_{\gamma_2} \eta_2)\,,
	\label{hsm}
\end{equation}
where the four mixing angles $\epsilon, \alpha, \gamma_1, \gamma_2$ are taken as the input parameters
in the scan of the scalar sector. This choice was instrumental for the so-called inversion algorithm 
developed in \cite{Liu:2024aew} for a much more efficient scalar space scanning procedure than the early study \cite{Ferreira:2017tvy}.

The key angle here is $\epsilon$, which plays a role similar to $\beta - \alpha$ of the 2HDM, namely, it controls the coupling of $\hsm$ to $ZZ$ and $WW$ pairs. The choice $\epsilon = 0$ would correspond to exact scalar alignment; but here, we consider a nonzero $\epsilon$. 
It also affects the couplings of $\hsm$ with the top quarks, both flavor-diagonal and flavor violating, such as $\hsm t\bar u$. 
The recent study \cite{Zhao:2025rds} showed that values $|\epsilon| > 0.4$
can be disregarded as they almost always lead to a conflict with top-quark-related measurements.

All these relations allow us to choose the following parameters as input when scanning the scalar sector:
\begin{equation}
	\left\lbrace v\,, \, \beta\,, \, \psi\,, \, \msm \right\rbrace , \quad
	\left\lbrace \epsilon\,, \, \alpha\,, \, \gamma_1\,, \, \gamma_2 \right\rbrace , \quad 
	\left\lbrace m_{H_1^\pm}^2\,, \, m_{H_2^\pm}^2\,, \, m_{11}^2 - m_{22}^2\,,\, \lambda_{89} \right\rbrace \,, 
\end{equation}
where $\lambda_{89} = \sqrt{\Im (\lambda_8)^2 + \Im (\lambda_9)^2 } > 0$, as \cite{Ferreira:2017tvy,Liu:2024aew}. 
By performing our scalar scan in these parameters, we keep the properties of $\hsm$ and charged Higgs bosons under tight control;
all other scalar quantities such as $\lambda_i$ or the properties of additional four neutral scalars can be numerically expressed through these input parameters. 

It must be added that since the model contains four extra neutral Higgs bosons and two charged scalars,
the full phenomenology goes beyond the decays of the SM-like Higgs.
Moreover, since the construct is based on a spontaneously broken global symmetry,
there is no SM-like limit in which Higgs bosons can be made arbitrarily heavy.
The recent detailed study of the scalar sector \cite{Liu:2024aew} showed that there is an upper limit 
of about 700~GeV on additional Higgs boson masses
and that one can find benchmark points in which all new Higgs bosons lie in the 300-700~GeV range.
However, an existence of several moderately heavy Higgs bosons, either neutral or charged, does not automatically run in conflict 
with negative searches for additional Higgs bosons at the LHC.
The main reason is that, with the misalignment parameter $\epsilon \ll 1$, the most visible production and decay channels
of the additional Higgs bosons are suppressed. For example, the recent study \cite{Zhao:2025rds}
demonstrated that even a Higgs boson lighter than the top quark an be easily made compatible 
with many top-quark-related observables.

The bottom line is that a comprehensive scan of the entire CP4 3HDM parameter space is definitely needed, 
and all available constraints on extra Higgs bosons must be taken into account. However, as mentioned in the introduction,
in order to make such a scan in the vast parameter space much faster, we should first select the most promising scenarios
in each sector of the model. This is what we do in the present work: we weigh upon the leptonic CP4-symmetric scenarios.
The outcome of the present study will be a key piece in the future comprehensive scan.

\subsection{The lepton Yukawa sector}

CP4 symmetry was extended to the quark Yukawa sector in \cite{Ivanov:2015mwl,Zhao:2023hws}, where the very specific forms of the quark Yukawa matrices were derived. These findings are equally applicable to the lepton Yukawa sector.
If needed, right-handed neutrinos can also be introduced in this model, as was done in \cite{Ivanov:2017bdx};
however, since our focus here is on the charged lepton signals, we do not introduce them in this work. 
To set up the notation, we write the lepton Yukawa sector in any 3HDM is given by
\begin{equation}
	-{\cal L}_Y = \bar{L}^0_L (\Gamma_1 \phi_1 + \Gamma_2 \phi_2 + \Gamma_3 \phi_3) e_R^0 + h.c.\label{Yukawa-general}
\end{equation}
The lepton generation indices are implicitly assumed. The superscript $0$ refers to the fact that these fields are the starting lepton fields before mass matrix diagonalization; we will remove the superscript when we pass to the physical leptons. When required to be invariant under CP4, the lepton Yukawa matrices $\Gamma_a, a=1,2,3$ can only assume very special textures labeled in \cite{Ferreira:2017tvy} as cases $A, B_1, B_2, B_3$, see Appendix~\ref{CP4-defs} for the explicit expressions in the case $B_1$.

It must be said that case $A$ corresponds to the trivial scenario of 
all leptons coupling only to the first Higgs doublet. As a result, all lepton fields transform trivially under CP4,
and LFV signals are absent. This possibility is certainly available within CP4 3HDM and was implicitly assumed in the previous works. 
What we explore in this paper is whether any {\em non-trivial} CP4 scenario, $B_1$, $B_2$, or $B_3$---with all $\Gamma_a$ unavoidably non-zero in each of these cases---can also be made compatible with the present LFV experimental constraints.

Upon electroweak symmetry breaking, the mass matrix of charged leptons in the real vevs basis of Eq.~\eqref{vevs1} is
\begin{equation}
	M_e^0 = \frac{v}{\sqrt{2}}(\Gamma_1 c_\beta + \Gamma_2 s_\beta c_\psi + \Gamma_3s_\beta s_\psi)\,.
\end{equation}
In general, this mass matrix is non-diagonal and complex, and can be diagonalized by the following three-dimensional unitary transformations of the lepton fields
$e_L^0 = V_{L} e_L$, $e_R^0 = V_{R} e_R$, $\nu_L^0 = U_L \nu_L$. 
These transformations lead to the Pontecorvo-Maki-Nakagawa-Sakata (PMNS) matrix: 
$\Vpmns = U_L^\dagger V_L$, or $ U_L = V_L \Vpmns^\dagger$.
The interaction of the neutral (complex) components of the scalar doublets with the leptons can be described in the Higgs basis as 
\begin{equation}
	\Gamma_1 \phi_1 + \Gamma_2 \phi_2 + \Gamma_3 \phi_3 = 
	\frac{\sqrt{2}}{v} (\Phi_1 M_e^0 + \Phi_2 N_{2}^0 + \Phi_3 N_{3}^0)\,,
\end{equation}
where
\begin{equation}
	N_{2}^0 = \frac{v}{\sqrt{2}}(-\Gamma_2 s_\psi + \Gamma_3 c_\psi)\,,\quad
	N_{3}^0 = \frac{v}{\sqrt{2} s_\beta} \Gamma_1  - M_d^0 \cot\beta  
	=  M_d^0 \tan\beta- \frac{v}{\sqrt{2}c_\beta}(\Gamma_2 c_\psi + \Gamma_3 s_\psi)\,.
	\label{N0_23}
\end{equation}
After the rotation of lepton fields, we get
$V_{L}^\dagger M_e^0 V_{R}  = {\rm diag} (m_e, m_\mu, m_\tau) \equiv  M_e$, 
$V_{L}^\dagger N_{2,3}^0 V_{R} = N_{2,3}$, and consequently,
$U_{L}^\dagger N_{2,3}^0 V_{R} = \Vpmns N_{2,3}$.
The Yukawa matrices $N_{2}$, $N_{3}$ describe the couplings across the lepton generations
of the complex fields present in $\Phi_2$ and $\Phi_3$, both neutral and charged. 
As for PMNS matrix parameters, we adopt the standard PDG values \cite{ParticleDataGroup:2024cfk, DayaBay:2024hrv, KM3NeT:2024ecf, Super-Kamiokande:2023jbt}:
\begin{equation}
	\theta_{12} = 33.6^\circ\,, \quad \theta_{23} = 46.9^\circ\,, \quad \theta_{13} = 8.5^\circ\,.
\end{equation}
As for the leptonic $CP$-violating phase, we choose $\delta = 210^\circ $ \cite{Super-Kamiokande:2023ahc}. 
Its experimental value still bears large uncertainties but it does not have any noticeable impact on our results;
this was verified by comparing the scatter plots presented below for the values of $\delta = 0, \pi, 3\pi/2$.

When performing a scan in the lepton Yukawa sector, we adapt the inversion procedure developed in \cite{Zhao:2023hws} for the quark sector.
We take the physical lepton masses and the PMNS matrix as input, parametrize $V_L, V_R$ appropriately, and express $N_2, N_3$ in terms of these parameters. Explicit expressions for $N_{2}$, $N_{3}$ in cases $B_1, B_2, B_3$ are provided in Appendix~\ref{expression-N2-N3}. 
It is worth pointing out that these coupling matrices are independent of the scalar sector angle $\psi$.

\subsection{Physical Higgs-lepton couplings}

In the Higgs basis of Eq.~\eqref{expansion-Higgs-basis}, the scalar sector contains five real and two complex scalar fields, apart from the would-be Goldstone bosons. Their couplings to the physical leptons are described by 
\begin{eqnarray}
	-{\cal L}_Y &=& \frac{1}{v}\bar{e}_{Li}\left[(M_e)_{ij}\, \rho_1 + (N_2)_{ij}\,(\rho_2 + i \eta_2) + (N_3)_{ij}\,(\rho_3 + i \eta_3)\right] e_{Rj} + h.c. \\
	-{\cal L}_Y' &=&  \frac{\sqrt{2}}{v}\bar{\nu}_{Li}\left[ (V_{\text{{\tiny PMNS}}} N_2)_{ij}\,w_2^+ 
	+ (V_{\text{{\tiny PMNS}}} N_3)_{ij}\,w_3^+\right] e_{Rj} + h.c. 
\end{eqnarray}
The physical neutral Higgs bosons $H_k$, with $k = 1, \dots, 5$, are linear combinations of the five neutral scalar fields, with their mixing described by a $5\times 5$ rotation matrix. Similarly, physical charged Higgs bosons $H_k^\pm$, $k=1,2$, arise from mixing of $w_2^\pm$ and $w_3^\pm$. All these expressions can be computed numerically for any point of the scan.

We parameterize the Yukawa interaction between physical leptons and physical scalars via their coupling matrices $Y^k$ (neutral Higgs bosons) and $ Z^k$ (charged Higgs bosons): 
\begin{eqnarray}
	-{\cal L}_Y &=& \bar{e}_{Li} Y^{k}_{ij} H_k \, e_{Rj} + \bar{e}_{Ri} (Y^{k\dagger})_{ij} H_k \, e_{Lj}\,. \\
	-{\cal L}_Y' &=& \bar{\nu}_{Li} Z^k_{ij} H^+_k e_{Rj} + \bar{e}_{Ri} (Z^{k\dagger})_{ij} H^-_k \nu_{Lj}\,.
\label{Yij}
\end{eqnarray}
For the SM-like Higgs boson $\hsm$, the coupling matrix, which we denote simply as $Y_{ij}$, can be written in terms of the familiar angles $\epsilon, \alpha, \gamma_1, \gamma_2$ defined in Eq.~\eqref{hsm}:
\begin{equation}
	Y_{ij} =c_\epsilon \cdot \frac{m_i}{v}\delta_{ij}  + s_\epsilon \cdot  \left[  \frac{(N_2)_{ij}}{v} ( c_\alpha c_{\gamma_1} + i\,  s_\alpha  s_{\gamma_2}) + \frac{(N_3)_{ij}}{v}(c_\alpha s_{\gamma_1} + i\, s_\alpha c_{\gamma_2})\right] \,.
	\label{Yij for hsm}
\end{equation}
The matrices $N_2$ and $N_3$ both modify the diagonal entries
and generate nonzero off-diagonal entries leading to LFV decays of the $125\GeV$ Higgs boson.

Each additional neutral Higgs boson $H_k$ can be expanded in the basis of $\Psi_{k'} \equiv (\rho_1, \rho_2, \eta_2, \rho_3, \eta_3)$
as $H_k = c^k_{k'}\Psi_{k'}$, where $c^k_{k'}$ are just the entries of the unitary $5\times 5$ diagonalizing matrix. 
Then, the coupling matrix $Y^k$ is 
\begin{equation}
	Y_{ij}^k =c_1^k \cdot \frac{m_i}{v}\delta_{ij}  +   \frac{(N_2)_{ij}}{v} ( c_2^k + i\,  c_3^k ) + \frac{(N_3)_{ij}}{v}(c_4^k + i\, c_5^k ) \,.
	\label{Yij for H_k}
\end{equation}
Finally, for the charged Higgs bosons $H_k^{\pm} = e^k_{k'} w_{k'}^\pm$, their coupling matrices $Z_{ij}^k$ are similarly written as 
\begin{eqnarray}
	Z_{ij}^k = \frac{\sqrt{2}}{v} \left[(\Vpmns N_2)_{ij} \cdot e_1^k + (\Vpmns N_3)_{ij} \cdot e_2^k \right]\,.
	\label{Yij for H+}
\end{eqnarray}

With this parametrization, and adopting the ultrarelativistic approximation for the produced leptons, the decay widths of any Higgs boson to two leptons can be written as
\begin{equation}
	\Gamma(H_k \to \ell_i^\mp \ell_j^\pm) = \frac{m_{H_k}}{8\pi}(|Y^k_{ij}|^2 + |Y^k_{ji}|^2)\,, \quad
	\Gamma(H_k \to \ell_i^- \ell_i^+) = \frac{m_{H_k}}{8\pi}|Y^k_{ii}|^2\,.\label{decay-widths}
\end{equation}
A key goal of this work is to identify the parameter regions where these Higgs-lepton couplings comply with experimental constraints on $\hsm$ lepton decays.

\subsection{Muon radiative decay: $\mu \to e \gamma$ at one loop}\label{subsection-mu-egamma}

\begin{figure}[h]
	\centering
	\includegraphics[width=0.9\linewidth]{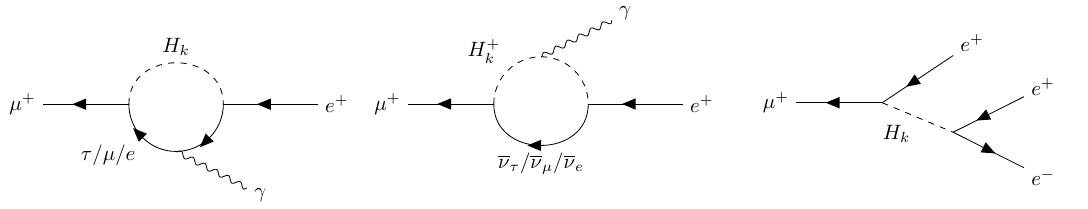}
	\caption{One-loop Feynman diagrams for $\mu^+ \to e^+\gamma$ due to neutral Higgs bosons (left) and charged Higgs bosons (middle).
	The right diagrams shows the tree-level decay of $\mu \to 3e$.}
	\label{one_loop}
\end{figure}
Another critical test of a non-trivially transforming leptonic sector of the CP4 3HDM is the radiative LFV decay $\mu \rightarrow e \gamma$. In NHDM, contributions to this decay start from one-loop diagrams shown in Fig.~\ref{one_loop}. 
For the one-loop result of this branching ratio, we adopt the expressions given in \cite{Crivellin:2013wna}:
\begin{eqnarray}
	\Br(\mu \rightarrow e \gamma) \equiv \mueg = \frac{m_\mu^5}{4\pi \Gamma_\mu}\left( |c_R^{e\mu}|^2 + |c_L^{e\mu} |^2  \right)\,.
	\label{mteg_formul}
\end{eqnarray}
The Wilson coefficients $c_R$ and $c_L$ receive contributions from both diagrams in Fig.~\ref{one_loop}:
$c_R^{e\mu} = (c_{R}^{e\mu})_{\textrm\tiny n.} + (c_{R}^{e\mu})_{\textrm\tiny ch.}$
and $c_L^{e\mu} = (c_{L}^{e\mu})_{\textrm\tiny n.} + (c_{L}^{e\mu})_{\textrm\tiny ch.}$.
All neutral Higgs bosons $H_k$ give the following contribution to $c_R^{e\mu}$:
\begin{eqnarray} 
	(c_{R}^{e\mu})_{\textrm\tiny n.} = \sum_{k=1}^{5}\sum_{j=e, \mu, \tau}\frac{-e}{192\pi^2 m_{H_k}^2} 
	\left[ Y^k_{ej} Y^{k*}_{\mu j} + \frac{m_e}{m_\mu} Y^{k*}_{je} Y^k_{j\mu} 
	- \frac{m_j}{m_\mu} Y^k_{ej}Y^k_{j\mu}(9-6\ln{\frac{m_{H_k}^2}{m_j^2}}) \right].\label{cRneutral}
\end{eqnarray}	 
The total neutral Higgs boson contribution $(c_{L}^{e\mu})_{\textrm\tiny n.}$ can be obtained from Eq.~\eqref{cRneutral} 
by exchanging $Y_{ij}$ with $Y_{ji}^*$. 
The total contribution of the two charged Higgs bosons $c_R^{e\mu}$ and $c_L^{e\mu}$ are given by 
\begin{eqnarray}
	(c_{L}^{e\mu})_{\textrm\tiny ch.} = \sum_{k=1}^2 \sum_{j=\nu_{e},\nu_\mu,\nu_\tau} \frac{e}{384 \pi^2 m_{H_k^{\pm}}^2} Z_{j\mu}^{k} Z_{je}^{k*} , \qquad
	(c_{R}^{e\mu})_{\textrm\tiny ch.} = \frac{m_e}{m_\mu} (c_{L}^{e\mu})_{\textrm\tiny ch.}\,.
\end{eqnarray}

Although the exact numerical value of the $\mu \to \gamma$ decay width depends in an intricate way on the interference among all terms 
and all lepton and Higgs boson contributions, these expressions already provide insights on which terms are expected to dominate.
These are the muon and tau-lepton contributions coming from the last term in Eq.~\eqref{cRneutral} with a large numerical factor.
Specifically, for muon, we find the product of a diagonal and an off-diagonal couplings $Y^k_{e\mu}Y^k_{\mu\mu}$
times a factor ${\cal O}(100)$. 
For the tau-lepton, the factor in the brackets is ${\cal O}(50)$ but it is further enhanced by $m_\tau/m_\mu \approx 17$.
However, it stands in front of two off-diagonal elements $Y^k_{e\tau}Y^k_{\tau\mu}$, which are controlled by different couplings.
Thus, we can expect that in a generic scan, the tau-lepton contribution dominates but we have enough freedom to
effectively switch it off by suppressing the $\tau$ coupling to lighter leptons.

The one-loop result is not the end of the story.
Back in the late 1980s, it was pointed out \cite{Weinberg:1989dx,Barr:1990vd,Dicus:1989va} that the two-loop diagrams could also give a sizable contribution if a large hierarchy between the various Yukawa couplings takes place. In particular, the recent Ref.~\cite{Davidson:2016utf} contains a detailed discussion, within the Type III 2HDM in the decoupling limit, of the magnitude of the two-loop Barr-Zee diagram and its dominant contribution coming from the top-quark loop.
Using the expressions presented in \cite{Davidson:2016utf},
we extract the factors from this top-quark Barr-Zee diagram 
\begin{equation}
	 \frac{e\, \alpha}{6\pi^3 m_t m_\mu}\sum_k Y^k_{e\mu}Y^k_{tt}\cdot f_k\,,
\end{equation} 
where $f_k$ are numerical factors for each Higgs boson of the order not too different from one.
This expression is to be compared with Eq.~\eqref{cRneutral}.
Despite significant numerical uncertainties involved,
the parametric factors found here are enough to make the following point.
We see that the two-loop contribution is suppressed by $\alpha$ but this suppression is 
more than compensated by the small $m_\mu$ in the denominator compared to the Higgs boson masses.
The key point is that this two-loop contribution contains the product of a leptonic and a quark sector couplings.
However, these two sectors are governed by different structures and independent free parameters.
In particular, we have freedom to independently adjust the two sectors in such a way that 
the Higgs bosons with large $Y_{tt}$ have vanishing $Y_{e\mu}$ couplings and vice versa, the Higgs bosons with a sizable $Y_{e\mu}$
feature a vanishingly small top-antitop coupling. 
This latter feature is perfectly compatible with CP4 3HDM;
in fact, a previous study \cite{Zhao:2025rds} found benchmark points with such top-quark-phobic Higgs bosons.

The bottom line is that the actual impact of the two loop diagrams can be substantial for many parameter sets,
increasing the $\mu\to e \, \gamma$ decay rate by orders of magnitude.
However, it is also possible to adjust the quark and lepton Yukawa structures in such a cooperative way
as to strongly suppresses the product $Y^k_{e\mu}Y^k_{tt}$ even for individual Higgs bosons, let alone their interference.
In those cases, the one-loop result alone offers a reasonable order-of-magnitude estimate of the total decay rate.

This situation can be quantitatively clarified only through a full phenomenological scan of CP4 3HDM.
We stress, however, that such a comprehensive study goes beyond the scope of the present work. 
he purpose of our work is to get all ingredients ready for an intelligent future scan of the entire CP4 3HDM phenomenology.
So, our goal is to determine which CP4 leptonic Yukawa sectors, if any, has a reasonable chance of satisfying LFV constraints.
If we find that, within a particular scenario, one-loop $\mu\to e \, \gamma$ prediction significantly exceeds the experimental bound,
we will deem this scenario ruled out. However is a scenario emerges in which
the one-loop prediction is {\em several orders} of magnitude below the experimental upper limit,
we label it as a potentially viable scenario, as it can survive even when two-loop diagrams are included.

It is with this logic in mind that we limit ourselves to the one-loop contribution to $\mu\to e\gamma$.

	\subsection{Decay $\mu \to 3e$}

The three-body decay $\mu^- \to e^-e^+e^-$ proceeds at tree level 
and involves contributions from all neutral Higgs bosons, see Fig.~\ref{one_loop}, right.
The expressions for the branching ratio can be found in \cite{Crivellin:2013wna}
and represented in a way similar to Eq.~\eqref{mteg_formul}:
\begin{equation}
	\mueee = \Br(\mu \to 3e) =\frac{m_\mu^5}{24(8\pi)^3 \Gamma_\mu} 
	\left( 
	2 \left| \sum_{k=1}^{5}\dfrac{Y_{\mu e}^{k*} Y_{ee}^{k}}{m_{H_k}^2} \right|^2 
	+ 2 \left|  \sum_{k=1}^{5}\dfrac{Y_{e\mu}^{k*} Y_{ee}^{k}}{m_{H_k}^2} \right|^2 
	+ \left|  \sum_{k=1}^{5}\dfrac{Y_{\mu e}^{k} Y_{ee}^{k}}{m_{H_k}^2} \right|^2 
	+ \left|  \sum_{k=1}^{5}\dfrac{Y_{e\mu}^{k} Y_{ee}^{k}}{m_{H_k}^2} \right|^2 
	\right) \,.\label{Brmueee}
\end{equation}
To gain some insight on its expected magnitude, we compare it with $\mu\to e \, \gamma$.
Both amplitudes contain products of two leptonic couplings, which can be of the same order,
divided by the Higgs boson mass squared.
The small numerical prefactors are of different origin: in the $\mu\to e \, \gamma$, it comes from the 
loop integration, while in $\mu \to 3e$, in emerges from the three-body phase space.
The $\mu\to e \, \gamma$ amplitude decay contains the electric charge $e \approx 0.3$
but is enhanced by a much larger logarithmic factor, which is absent for the tree-level $\mu \to 3e$ process.
Although the summations over the five Higgs bosons are not exactly the same,
we generically expect $\mu\to e \, \gamma$, with with the $\tau$-loop suppressed, to exceed $\mu \to 3e$.
Still, to make sure we control this process, we implement in our numerical scan.

\section{Numerical results}\label{section-numerical}

\subsection{The scan procedure and constraints}

With the expressions for the leptonic decays of $\hsm$ and for the branching ratio of $\mu \to e\gamma$, we are now ready to perform a numerical scan in the CP4 3HDM parameter space and explore the constraints on the model arising from the experimental data. 
In this section, we present the results of this study, using as the key example the CP4-invariant lepton sector $B_1$. 
As we will see below, this is the most promising scenario, as its parameter space includes regions that satisfy all the experimental bounds 
by a large margin. The two other cases, $B_2$ and $B_3$, can be treated in a similar way and typically lead to stronger LFV signals; 
their results are reported at the end of this section. 

Our scan procedure consists of two steps. First, we scan the scalar sector and filter out a substantial sample of scalar parameter space points that satisfy all the theoretical constraints; then, this sample is used as input for the lepton sector scan. It is known that a generic scan in the scalar sector of the CP4 3HDM typically produces one or several Higgs bosons lighter than the top quark \cite{Ferreira:2017tvy}.
Although such situations are not ruled out \cite{Zhao:2025rds}, we prefer to focus on the region with all non-standard scalars as heavy as possible.
To this end, we follow the procedure developed in \cite{Liu:2024aew} and perform a restricted scan in the scalar sector that highlights the high-mass region. This is done with the following ranges of the input parameters:
\begin{eqnarray}
	&&\tan\beta \in [0.5, 2]\,, \quad \tan\psi \in [0.5, 3]\,, \quad 
	|\epsilon| < 0.3\,, \quad |\tan\alpha| < 0.05\,, \quad	|\tan\gamma_{1}|, \ |\tan\gamma_{2}| < 0.3\,,\label{focused-scan-range}\\[1mm]
	&& 
	m_{11}^2 - m_{22}^2 \in [-(300\,\GeV)^2, (300\,\GeV)^2]\,, \quad m_{H_{k}^\pm}^2 \in [0, (700\,\mbox{GeV})^2]\,,
	\quad \lambda_{89} \in (0, 4)\,.
	\nonumber
\end{eqnarray}
Then we reconstruct numerically all the parameters of the potential 
and impose the standard theoretical constraints: the bounded-from-below (BFB) conditions \cite{Ivanov:2015mwl}; the unitarity and perturbativity constraints following the simplified approach of \cite{Ferreira:2017tvy} (although it must be mentioned that the exact conditions are now known \cite{Bento:2022vsb}); the electroweak precision parameters $S, T$, and $U$, which are required to lie within the $2\sigma$ ranges of their experimental values \cite{Grimus:2008nb}. We then compute the scalar mass matrices, diagonalize them numerically,
and obtain the masses of all other Higgs bosons, as well as the entries of the diagonalizing matrices $c^k_{k'}$ and $e^k_{k'}$,
and select only the points for which all additional neutral Higgs bosons are heavier than $\hsm$. 

As for the leptonic sector scan, we adopt the inversion procedure developed in \cite{Zhao:2023hws} for the lepton sector,
using the scalar sector sample as input, along with the angles of the rotation matrices $V_L$, $V_R$, $U_L$,
which are constrained by the PMNS matrix.
In the initial scan, we allow these angles to vary over their entire domains. However, we will later switch to a restricted scan where
the relevant angles are sampled in a tightly constrained interval to further suppress LFV signals.

For the experimental limits on the LFV decay branching ratio of $\hsm$, we use the 95\%CL limits listed by the Particle Data Group (PDG) \cite{ParticleDataGroup:2024cfk}, which are based on the experimental results reported in \cite{CMS:2023pte,ATLAS:2023mvd,CMS:2021rsq}:
\begin{equation}
	\Br(\hsm\to e\mu) < 4.4\times 10^{-5}, \quad 
	\Br(\hsm\to e\tau) < 2.0\times 10^{-3}, \quad 
	\Br(\hsm\to \mu\tau) < 1.5\times 10^{-3}\,.
	\label{Higgs-LFV-PDG}
\end{equation}
For the lepton flavor-conserving (LFC) decay modes, we use the experimental results from \cite{CMS:2022urr, ATLAS:2025coj, ATLAS:2022vkf, CMS:2022dwd} compiled by the Particle Data Group:
\begin{equation}
	\begin{tabular}[t]{lcccc}
		\toprule
		&\qquad & SM, Ref.~\cite{LHCHiggsCrossSectionWorkingGroup:2016ypw} &\qquad& PDG \\
		\midrule %
		$\Br(\hsm\to ee)$ && $5\times 10^{-9}$ && $< 3.0 \times 10^{-4}$  \\[1mm] 
		$\Br(\hsm\to \mu\mu)$ && $2.15\times 10^{-4}$ && $\mu_{\mu\mu}^h = 1.4 \pm 0.4$  \\[1mm]
		$\Br(\hsm\to \tau\tau)$ && $6.2\times 10^{-2}$ && $\mu_{\tau\tau}^h = 0.97 \pm 0.11$  \\
		\bottomrule
	\end{tabular}  
	\label{table-LFC}
\end{equation}
Here, $\mu_{\ell\ell}^h = \Br(\hsm\to \ell\ell)/[\Br(\hsm\to \ell\ell)]_{\rm SM}$ are the usual signal strengths of the leptonic decays channels.

Finally, the experimental upper limit is $\Br(\mu \rightarrow e\gamma) < 1.5\times 10^{-13}$, \cite{MEGII:2025gzr},
while the upper bound for $\mueee$ is $1.0 \times 10^{-12} $ \cite{SINDRUM:1987nra}.

\subsection{Case $B_1$: leptonic decays of $\hsm$}

\begin{figure}[h]
	\centering
	\includegraphics[width=0.3\linewidth]{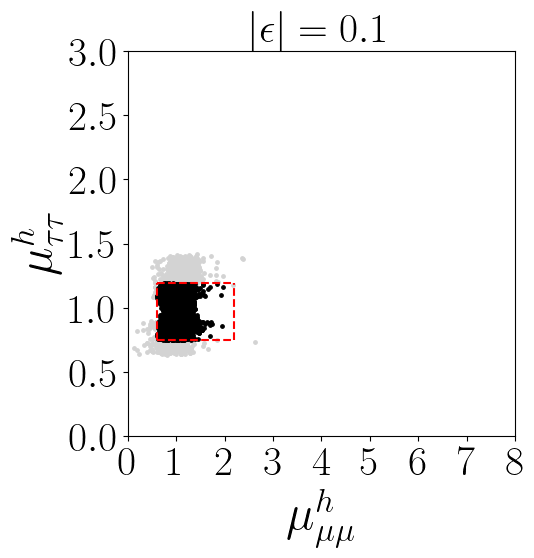}\quad
	\includegraphics[width=0.3\linewidth]{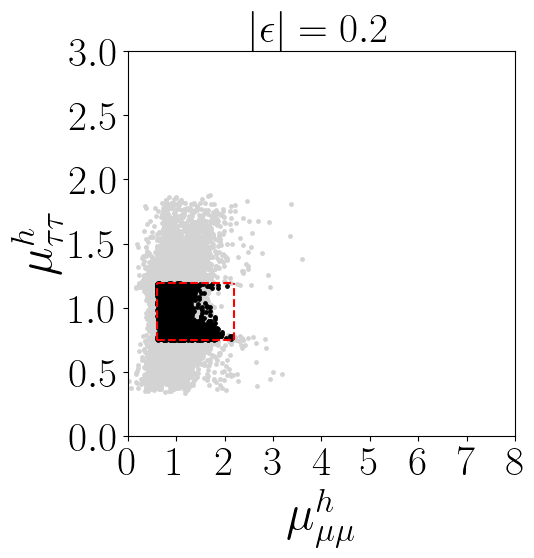}\quad
	\includegraphics[width=0.3\linewidth]{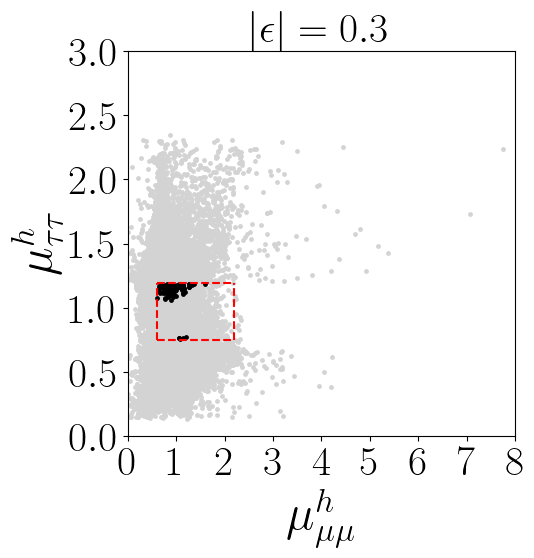}
	
	\includegraphics[width=0.3\linewidth]{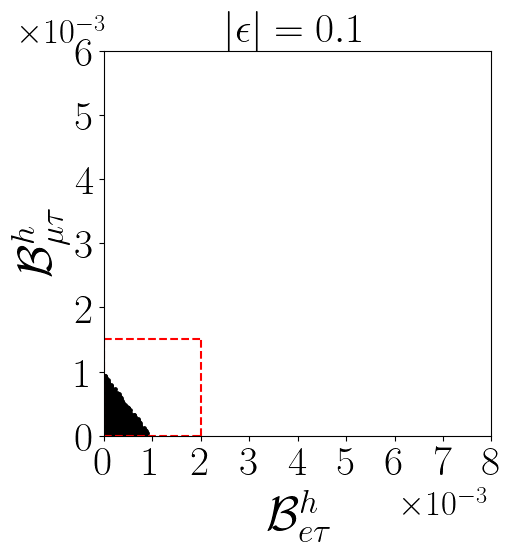}\quad
	\includegraphics[width=0.3\linewidth]{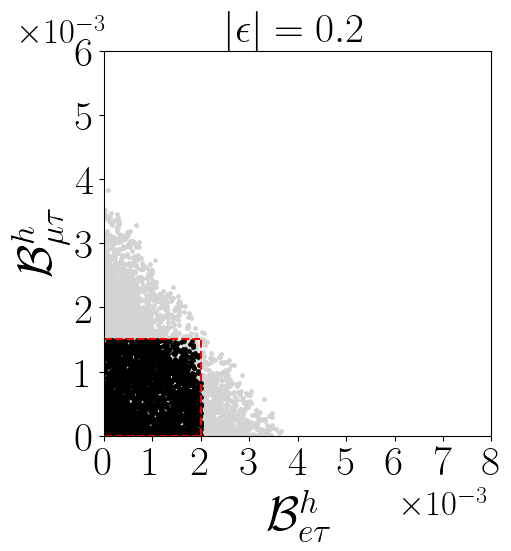}\quad
	\includegraphics[width=0.3\linewidth]{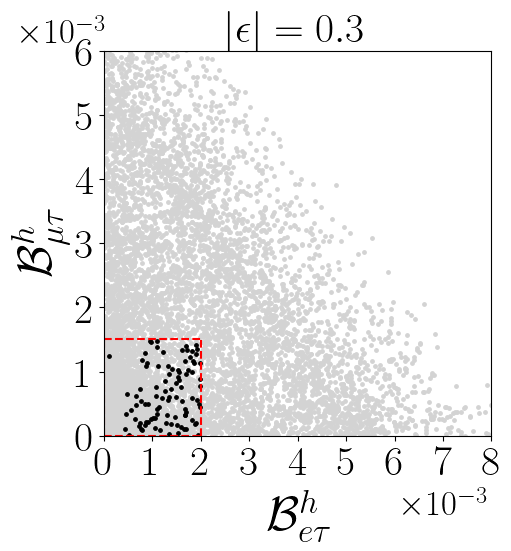}
	\caption{The impact of the misalignment angle $\epsilon$ on the LFC (top row) and LFV (bottom row) decays of $\hsm$. The dashed red box shows the experimental limits. The points which satisfy all six experimental limits are shown as black dots.}
	\label{hsm-ce}
\end{figure}

Before presenting the results of the full scan, let us explore how the leptonic decays of $\hsm$ depend on the parameters.
As can be seen from Eq.~\eqref{Yij for hsm}, the key role is played by the misalignment angle $\epsilon$: the close this angle to zero,
the weaker are the departures from the SM predictions, both in the LFC and LFV decays.
For small $\epsilon$, the branching ratios of the LFC and LFV decays deviate from their SM values by $s_\epsilon$ and $s_\epsilon^2$, respectively.
The patterns of these departures, which come from the matrices $N_2$ and $N_3$, significantly depend on $\tan\beta$ but not on $\psi$. 

In Fig.~\ref{hsm-ce}, we show the impact of $\epsilon$ on the LFC (top row) and LFV (bottom row) decays of $\hsm$.
In these plots, we use the shorthand notation $\mathcal{B}_{\ell_i \ell_j}^h \equiv \Br(\hsm\rightarrow \ell_i \ell_j)$.
When a generated point satisfies all six experimental limits, we plot it as a black dot;
if at least one constraint is violated, it is plotted in gray. The red dashed boxes indicate the experimental constraints within each plane.
As $|\epsilon|$ increases, the departures from the SM predictions grow as expected. For $|\epsilon| \le 0.1$, most sampled points
are still within the experimental limits. For $|\epsilon| > 0.3$, very few randomly selected points fall within the experimental limits,
the strongest constraints coming from $\hsm \to \tau\tau$ and the LFV decays. In addition, it was shown in \cite{Zhao:2025rds}
that the couplings $\hsm t\bar t$ and $\hsm u \bar{t}$ become strong and violate experimental constraints 
on the Higgs-top processes. With these results, we focus on the region $|\epsilon| \le 0.3$ in the subsequent analysis.
Note that these constraints are stronger than those coming from the $\hsm VV$ couplings \cite{Zhao:2025rds}.

\begin{figure}[h]
	\centering
		\includegraphics[width=0.3\linewidth]{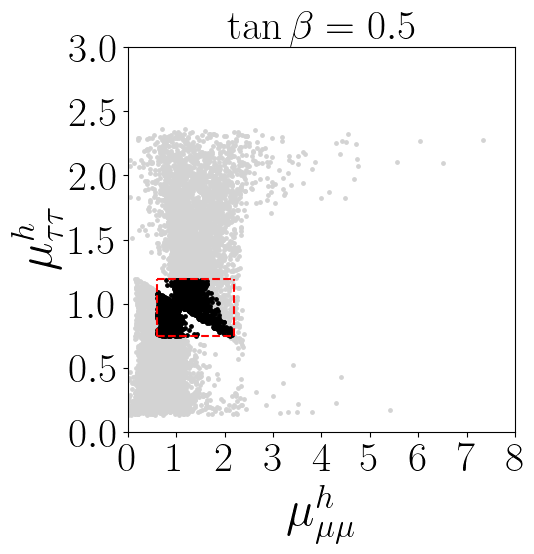}\quad
		\includegraphics[width=0.3\linewidth]{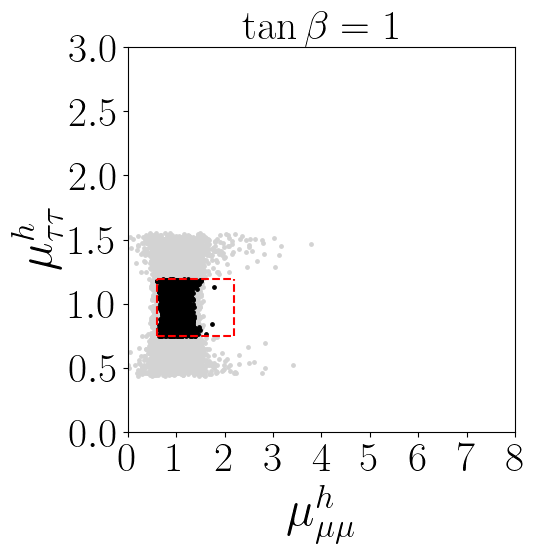}\quad
		\includegraphics[width=0.3\linewidth]{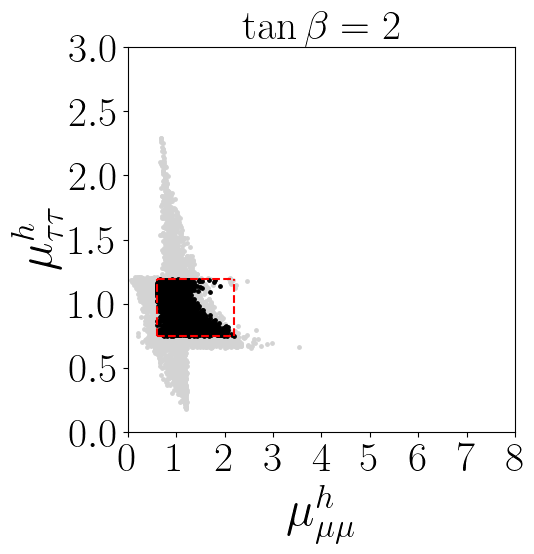}

		\includegraphics[width=0.3\linewidth]{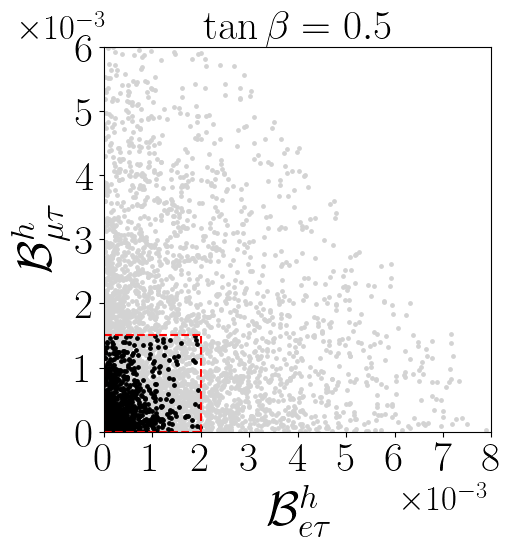}\quad
		\includegraphics[width=0.3\linewidth]{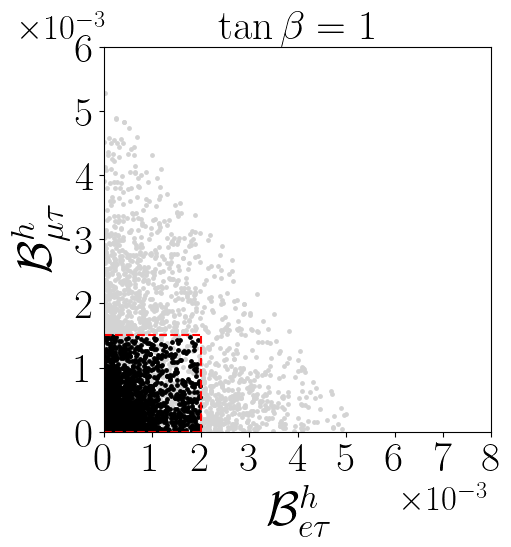}\quad
		\includegraphics[width=0.3\linewidth]{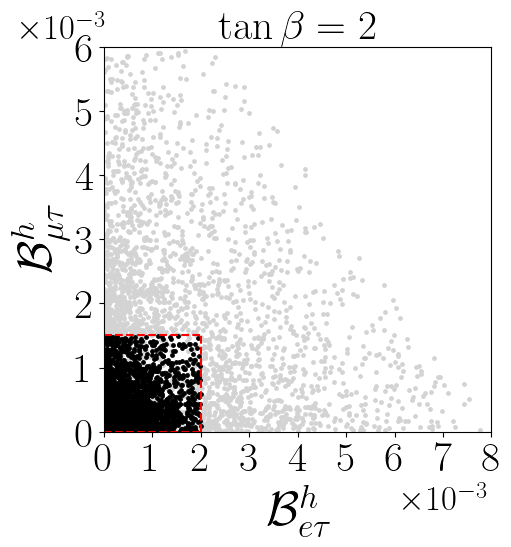}
	\caption{The impact of $\tan\beta$ on the LFC and LFV decays of $\hsm$. The color conventions are the same as in Fig.~\ref{hsm-ce}.}
	\label{hsm-tb}
\end{figure}

Another important parameter, $\tan\beta$, influences the patterns of the Yukawa matrices $N_2$ and $N_3$. Ref.~\cite{Zhao:2023hws}
contains explicit expression for $N_2, N_3$ for up and down quarks. One leptonic sector matrices
repeat their expressions for the down-quark sector and are collected in Appendix~\ref{expression-N2-N3}. 
For the Yukawa sector $B_1$, which our main scenario here, the matrix $N_2$ has the form
\begin{eqnarray}
	N_{2} = V_{L}^\dagger\, R_3^0 \, V_{L}\cdot M_d\,, \quad \text{where} \quad R_3^0 = {\rm diag}(\cot\beta, \cot\beta, -\tan\beta)\,.
\end{eqnarray}
It is clear from this expression that the magnitude of the LFV elements is controlled by the rotation angles of the left-handed charged lepton diagonalizing matrix $V_L$,
while its diagonal elements are controlled by $\tan\beta$.
As for $N_3$, its main dependence on $\beta$ comes from the $1/\sin\beta$ prefactor, see Appendix~\ref{expression-N2-N3}.
However, the physical Higgs bosons receive contributions from both $N_2$ and $N_3$, which results in a non-trivial interplay of the two structures.

Fig.~\ref{hsm-tb} shows the LFC and LFV decays for $\tan\beta = 0.5$, 1, and 2; the other parameters are scanned over. 
Both LFV decay contours/density and LFC decay contours exhibit strong correlation with $\tan\beta$. 
Overall, within the selected scanning range of $\tan\beta$, a significant fraction of random scan points satisfy all the six experimental constraints.

\begin{figure}[h]
	\centering
	\includegraphics[width=0.36\linewidth]{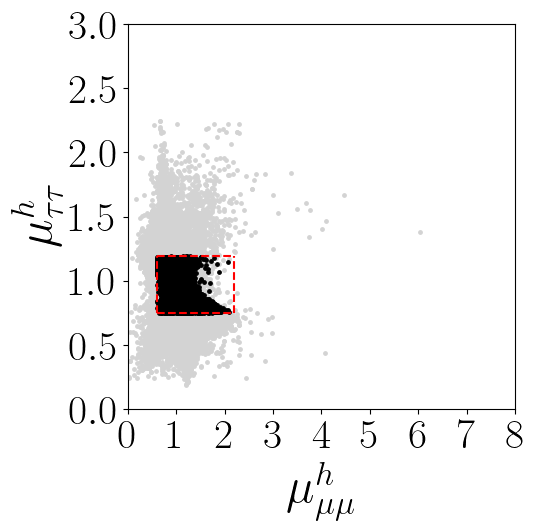}\qquad
	\includegraphics[width=0.35\linewidth]{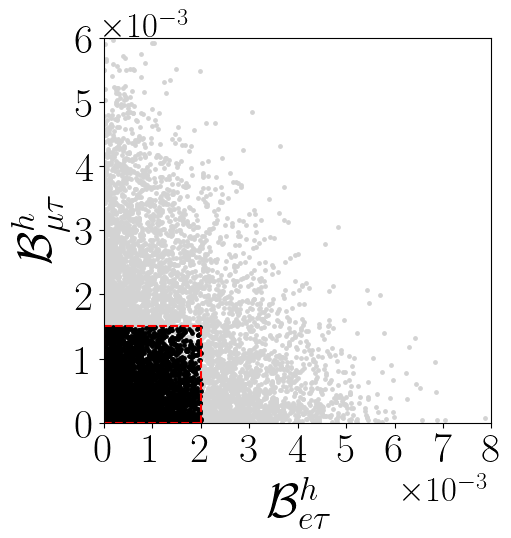}
	\caption{A comprehensive numerical scan of $\hsm$ lepton decay in high-mass region.}
	\label{higgs-decay}
\end{figure}

With these insights in mind, we can combine the scalar sector scan within the high-mass range Eq.~\eqref{focused-scan-range} 
with the leptonic Yukawa sector and explore the leptonic decays of $\hsm$. 
These results are summarized in Fig.~\ref{higgs-decay}. The difference with respect to the previous plots is that we now do not slice the parameter space
at fixed $\epsilon$ or fixed $\tan\beta$ but scan over all parameters. We see that such a scan easily produces many parameter space points that meet all six $\hsm$ experimental constraints. 
Thus, it confirms the model's compatibility with $\hsm$ leptonic decay data and allows us to continue exploration. 
In the subsequent discussion, we will only keep the black points, which satisfy all six leptonic decay constraints.

\subsection{Case $B_1$: decays $\mu\rightarrow e \, \gamma$ and $\mu\to3e$}

\begin{figure}[h]
		\centering
		\includegraphics[width=0.43\linewidth]{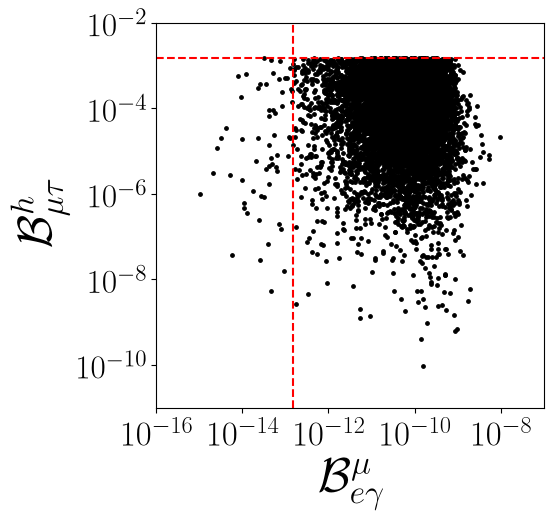}\qquad
		\includegraphics[width=0.4\linewidth]{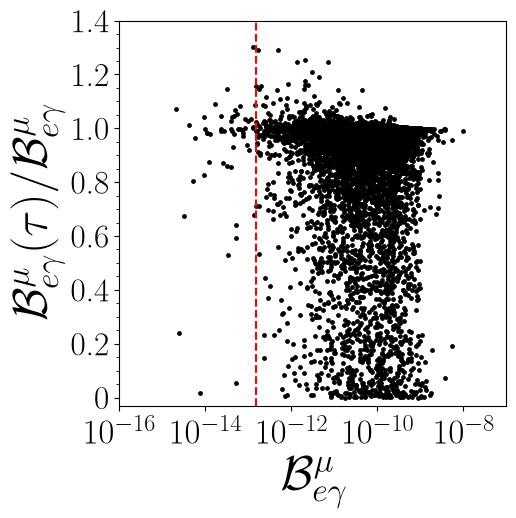}
	\caption{Left: the results of the full Yukawa scan in the plane $\hmutau$ vs. $\mueg$. Right: the distribution of the ratio $\mueg (\tau)/\mueg$, the contribution of $\tau$ propagator diagram to $\mu\rightarrow e\gamma$. The red dashed lines show the experimental upper limits.}
	\label{mteg_full}
\end{figure}

Next, we turn to $\mu\rightarrow e \, \gamma$, which receives contributions from extra Higgs bosons, both neutral and charged. 
We first perform a numerical scan in a similar way as above: the scalar sector parameters are scanned within the ranges given in Eq.~\eqref{focused-scan-range},
while for the leptonic sector scan, we use uniform sampling of all the angles of the matrices $V_L$, $V_R$, see Appendix~\ref{appendix-U3} for the definitions. 
We refer to this scan as the full Yukawa scan and show its outcome in Fig.~\ref{mteg_full}. 

The key observable is $\mathcal{B}^\mu_{e\gamma} \equiv \Br(\mu \rightarrow e\gamma)$ shown on the horizontal axis on both plots; the vertical red dashed line indicates its
MEG II upper limit $1.5\times 10^{-13}$ taken from \cite{MEGII:2025gzr}. 
The vertical axis on Fig.~\ref{mteg_full}, left, shows one of the most constraining LFV Higgs decays, $\hsm\to\mu\tau$.
We see that almost all points significantly exceed the MEG II upper limit, with only very few points passing this constraint.

Fig.~\ref{mteg_full}, right, provides an additional insight into the dynamics of this decay. Here, we show the result of the $\mu\to e\gamma$ calculation
in which only the loops with the $\tau$ lepton is taken into account. We denote the result as $\mueg(\tau)$
and show on the vertical axis of Fig.~\ref{mteg_full}, right, its relative value to the full $\mueg$ calculation.
The plot confirms our initial expectation from the analytical expression Eq.~\eqref{cRneutral} 
of the $\tau$-loop dominance for majority of points.
Moreover, the closer this ratio is to one, the easier it is to further suppress the value of $\mueg$.

This insight suggests a strategy to further suppress $\mu\to e\gamma$: we need to sample the leptonic parameter space in such a way that the LFV couplings
of the $\tau$ and $\hsm$ are suppressed.
To this end, we follow the procedure developed in \cite{Zhao:2023hws,Zhao:2025rds} and perform a restricted Yukawa scan, in which we limit the angles as
\begin{eqnarray}
	\theta_{13}, \theta_{23} \in [-\theta_{max}, \theta_{max}], \quad \theta_{max}=\pi/100 \,.
\end{eqnarray}
Then, the rotation matrices $V_L$ and $V_R$ of nearly block diagonal form
\begin{equation}
	V_L, V_R  \sim 
	\begin{pmatrix}
		\times & \times & \cdot \\
		\times & \times & \cdot \\
		\cdot & \cdot & \times
	\end{pmatrix},
\end{equation}
where each $\cdot$ stands for a non-zero but small elements, while $\times$ denotes an element of order one. 
Under such approximation, $N_2, N_3$ simplify to
\begin{equation}
	N_2 \sim 
	\begin{pmatrix}
		\times & \cdot & \cdot \\
		\cdot & \times & \cdot \\
		\cdot & \cdot & \times
	\end{pmatrix}, \quad
	N_3 \sim 
	\begin{pmatrix}
		\times & \times & \cdot \\
		\times & \times & \cdot \\
		\cdot & \cdot & \cdot
	\end{pmatrix}\,,
\end{equation}
which automatically suppresses the contributions from the $\tau$-induced loop to $\mueg$. 
Moreover, this procedure automatically suppressed the LFV Higgs boson decays $\mathcal{B}_{e\tau}^h$ and $\mathcal{B}_{\mu\tau}^h$. 

\begin{figure}[!h]
		\centering
		\includegraphics[width=0.42\linewidth]{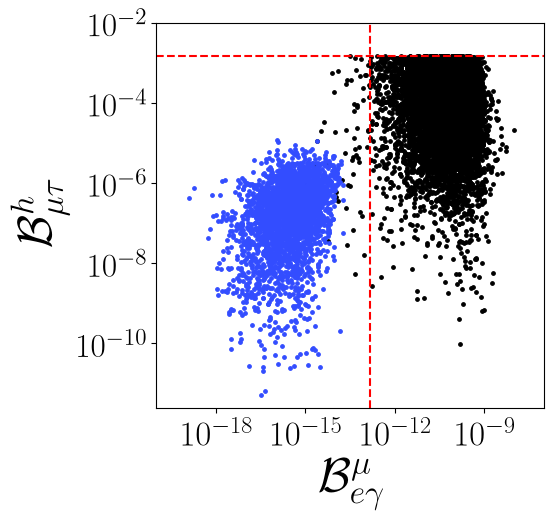}\qquad
		\includegraphics[width=0.42\linewidth]{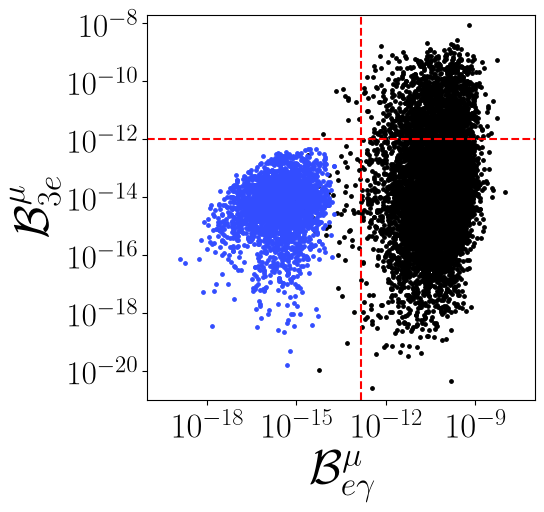}
	\caption{The distributions of $\mueg$ vs. $\hmutau$ (left) and $\mueee$ (right). 
		The black and blue points correspond to the full and restricted ($\theta_{max} = \pi/100$) Yukawa scans, respectively.}
	\label{mteg_full_res}
\end{figure}

With this discussion in mind, we performed the focused Yukawa scan and show its results with blue points in Fig.~\ref{mteg_full_res}, left,
which is to be compared with the black points that represent the outcome of the full Yukawa, the same as in Fig.~\ref{mteg_full}, left.
We observe a dramatic reduction in $\mueg$ and in the Higgs bosons LFV decays. 
With the restricted scan, almost all of the points easily pass the $\mueg$ constraints, often by several orders of magnitude. 
This leaves us a safe margin to expect that even with the two-loop contributions included,
many of these points still have decent chances to stay below the experimental limits.
We also checked the branching ratio of $\mu \to 3e$, both in the full and restricted scans. The results for $\mueee$ are
shown against $\mueg$ in Fig.~\ref{mteg_full_res}, right. We see that a restricted scan suppressed $\mu \to 3e$ as well.

We conclude that by implementing a restricted Yukawa scan, we can easily construct CP4-invariant leptonic cases $B_1$ that pass all the checks from $\hsm$ leptonic decays and from muon decays $\mu\rightarrow e\gamma$ and $\mu\to 3e$.

\subsection{Cases $B_2$ and $B_3$}

\begin{figure}[h]
	\centering
	\includegraphics[width=0.45\linewidth]{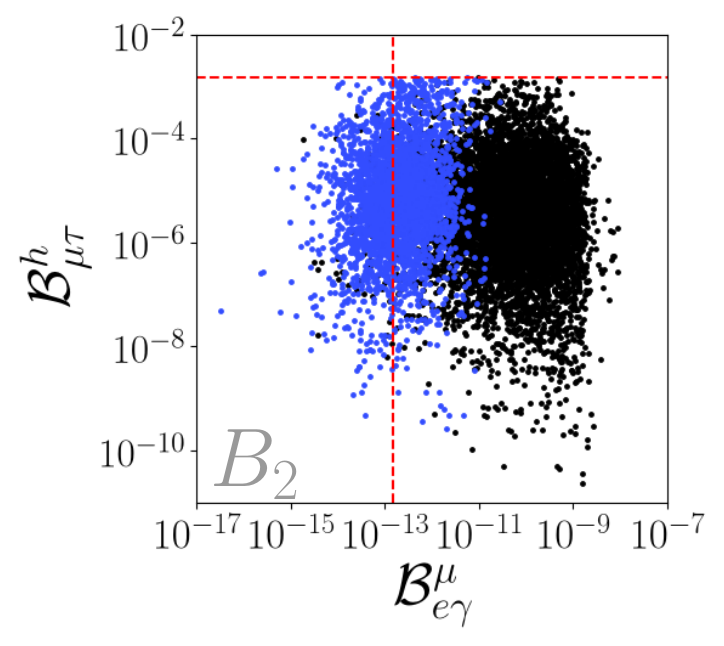}\quad
	\includegraphics[width=0.45\linewidth]{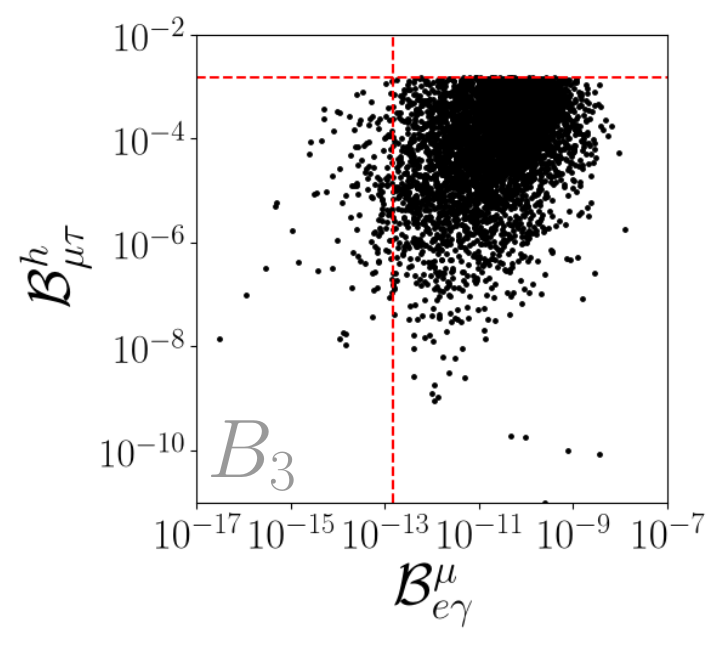}
	\caption{The full and restricted Yukawa scans for the Yukawa scenarios $B_2$ (left plot) and $B_3$ (right plot), with the same color code as in Fig.~\ref{mteg_full_res}. }
	\label{mteg-B2-B3}
\end{figure}

We also implemented and explored the two other CP4-invariant leptonic sectors, $B_2$ and $B_3$.
The analysis of the Higgs boson leptonic decays, both LFC and LFV, produced plots similar to Fig.~\ref{higgs-decay}.
Although the shapes of regions covered by gray points were somewhat different, we could still find many black points
that passed all six Higgs decay constraints. However, the results on $\mueg$ shown in Fig.~\ref{mteg-B2-B3} display a very different picture than in the case $B_1$.
As expected, the full Yukawa scan almost always violated the $\mueg$ upper limit.
However, for case $B_2$, we found a different version of the restricted Yukawa scan,
$\theta_{12}, \theta_{13} \in [-\theta_{max}, \theta_{max}]$, with $\theta_{max}=\pi/100$,
in the right-handed lepton rotation matrix $V_R$. 
This choice led to a significant reduction of the $\mueg$ decay rate. The results of this scan are shown by blue points in Fig.~\ref{mteg-B2-B3}, left.
Still, for most of the points, the suppression is insufficient. Although this scenario is not readily excluded, 
it appears very fine-tuned, with little margin remaining. 
In the light of potentially significant two-loop contributions, which further enhance this decay, we do not consider this scenario promising.

As for the scenario $B_3$, the special form of its Yukawa coupling patterns discussed in \cite{Zhao:2023hws} does not allow for any simple restriction on the rotation matrix angles that could help us significantly suppress $\mueg$. We consider this case to be nearly ruled out by the MEG II upper limit.
 
In short, we believe that case $B_1$ is the most promising candidate for a non-trivial extension of the CP4 symmetry to the leptonic sector.
We stress that we do not claim that all points deemed viable in our study will eventually satisfy all leptonic experimental constraints
once the quark sector is added and two-loop contributions to $\mu \to e\gamma$ are included. We only claim that 
only the $B_1$ CP4-symmetric scenario has good chances to provide such benchmark points.

\section{New leptonic signatures from additional Higgs bosons}\label{section-146}

The previous section dealt with the leptonic decays of the SM-like Higgs boson as well as the $\mu\to e\gamma$ decays. However, the multi-Higgs model we consider contains many more Higgs scalars, which can also couple to leptons and lead to leptonic signals at different invariant masses.
In the vicinity of the alignment limit, it is natural to expect that all these potential signals are suppressed. In fact, for $\epsilon = 0$,
such signals vanish altogether, at least at tree level.

However, it turns out that the LHC data actually do contain hints of new scalars decaying leptonically.
Indeed, the CMS collaboration reported in \cite{CMS:2023pte} an intriguing $e\mu$ excess for a new resonance at the invariant mass of $146 \GeV$,
with a local significance of $3.8 \sigma$ and a global significance of $2.8\sigma$. 
If the excess is real, the best fit for its signal strength was found by CMS to be
	\begin{equation}
		\sigma(pp\rightarrow X(146) \rightarrow e\mu) = 3.89^{+1.11}_{-1.08}\mbox{(stat)}{}^{+0.57}_{-0.34}\mbox{(syst)}\, \mbox{fb}.
		\label{CMS-excess}
	\end{equation}

Such a signal could be accommodated in many models, including ours. However, most models predict that an $H \to e\mu$ hint
should be accompanied by other leptonic signals at the same invariant mass, both LFC and LFV. 
The exact relation among these signals depends on the model and its free parameters.

	What makes CP4 3HDM different from most of these models is its built-in generation-mixing feature, see Appendix~\ref{CP4-defs}.
As the CP4 transformation acts in both scalar and Yukawa sectors by mixing generations,
the physics Higgs-lepton coupling matrices $Y^k_{ij}$ in Eq.~\eqref{Yij for H_k}
do not follow the traditional hierarchy typical in other models, such as for 2HDM. 
In particular, $Y^k_{ee}$ and $Y^k_{e\mu}$ can be as large as, or even larger than $Y^k_{\mu\mu}$.
This is why a hint of a new Higgs boson appearing first in the $e\mu$ channel is very much compatible
with the general expectations of CP4 3HDM.
Thus, it becomes an interesting question whether the predictions of the other channels are still compatible 
with the non-observation of the corresponding signals.

This is what we report in this section. We consider the CP4 3HDM points with the Yukawa scenario $B_1$ that pass all the constraints
discussed previously and select only those of them that contain a new neutral Higgs boson, denoted as $\hnew$, in the mass range 140--150~GeV,
the exact value 146~GeV being inessential in our study. To get a broad picture without unnecessary complications, we also require
that $\hnew$ is the only new Higgs boson in this mass range. 

\begin{figure}[h]
	\centering
	\includegraphics[width=0.4\linewidth]{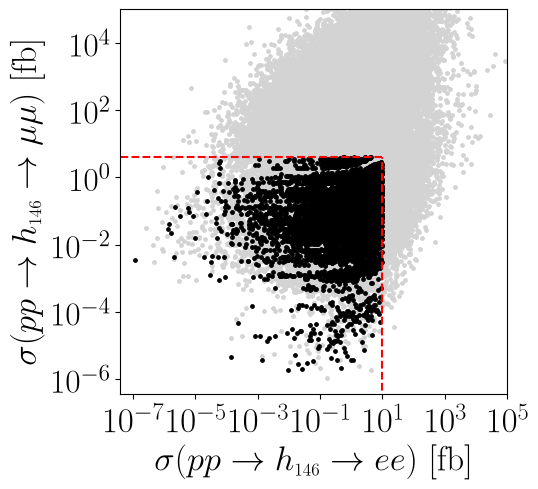}
	\caption{The distribution of the leptonic signals $\sigma(pp\rightarrow \hnew \rightarrow ee)$ and $\sigma(pp\rightarrow \hnew \rightarrow \mu\mu)$ 
		in the CP4 3HDM with scenario $B_1$ passing all the $\hsm$ and $\mu\to e\gamma$ constraints. 
		The points that satisfy all six new experimental limits for $\hnew$ are shown by black dots.}
	\label{146}
\end{figure}

Next, since we do not know the preferred production and decay channels for $\hnew$, we cannot predict its production cross section
and its decay branching ratios, as they depend on all the sectors of the model. However, we do not actually need that information. 
It suffices to rely on the signal reported by CMS, Eq.~\eqref{CMS-excess}, 
which we fix at $\sigma(pp\rightarrow \hnew \rightarrow e\mu) = 3.9$~fb, and define the other leptonic signals as
\begin{eqnarray}
	\sigma(pp\rightarrow \hnew \rightarrow \ell_i \ell_j) = \sigma(pp\rightarrow \hnew \rightarrow e\mu) \times 
	\frac{\Gamma (\hnew \rightarrow \ell_i \ell_j)}{\Gamma (\hnew \rightarrow e\mu)}\,.
\end{eqnarray}
Thu, we can predict all the leptonic signals coming from $\hnew$ decays by computing the ratios within our model and multiplying 
them by 3.9~fb.
These predictions must then be confronted with available experimental data.
Although the direct experimental upper limits for the $\hnew$ signal in various $\ell_i \ell_j$ channels are not available, 
the ATLAS and CMS collaborations published the results of their generic new scalar searches in various leptonic final states \cite{CMS:2022urr,ATLAS:2025coj,CMS:2022kdi, ATLAS:2023mvd, CMS:2021rsq}.
We studied the data point distributions in the invariant mass range of interest, compared them with
the $\hsm$ signal (if detected) and upper limits (if not seen), and knowing the SM-Higgs production cross section,
we conservatively placed the following constraints:
\begin{eqnarray}
	&&\sigma(pp\rightarrow \hnew \rightarrow ee) < 10 \, \text{fb}, \qquad \sigma(pp\rightarrow \hnew \rightarrow \mu\mu) < 4 \, \text{fb}, \qquad \sigma(pp\rightarrow \hnew \rightarrow \tau\tau) < 1000 \, \text{fb}, \nonumber\\
	&& \sigma(pp\rightarrow \hnew \rightarrow e\tau) < 100 \, \text{fb}, \qquad \sigma(pp\rightarrow \hnew \rightarrow \mu\tau) < 100 \, \text{fb}. \label{h146-bounds}
\end{eqnarray}
In Fig.~\ref{146}, we illustrate the results of this study with a representative plot 
of $\sigma(pp\rightarrow \hnew \rightarrow \mu\mu)$ vs. $\sigma(pp\rightarrow \hnew \rightarrow ee)$. 
We stress again that all points already satisfy the constrained discussed in the previous section.
Among them, black points correspond to the cases when all $\sigma(pp\rightarrow \hnew \rightarrow \ell_i \ell_j)$ signals
satisfy the experimental limits in Eq.~\eqref{h146-bounds}. The most salient feature of this plot is the sheer range of the values
that $\sigma(pp\rightarrow \hnew \rightarrow \mu\mu)$ and $\sigma(pp\rightarrow \hnew \rightarrow ee)$ can attain
even when $\sigma(pp\rightarrow \hnew \rightarrow e\mu)$ is fixed at 3.9~fb,
which highlights the peculiar couplings of a new Higgs boson across leptonic generations.
Many of the points are excluded by orders of magnitude, but there also remain many examples where the signals are well below the current upper limits.

	We stress again that the $ee$ channel is not always suppressed---let alone suppressed by $(m_e/m_\mu)^2$---with respect to the $\mu\mu$ channel.
In fact, typical are the situations where the $ee$ channel dominates over $\mu\mu$ and $e\mu$.
These features are quite distinct from the most multi-Higgs models such as 2HDM with LFV 
and are driven by the unavoidable generation-mixing property of CP4.

Although there is no smoking-gun observable in support of CP4 3HDM as compared to other models, 
this finding definitely has interesting implications for future experimental searches. The current collider searches of the additional Higgs bosons via leptonic channels focus on LFV channels like $e\mu$. However, our results show that the $ee$ decay channel of $\hnew$ could also be within the reach of the HL-LHC \cite{Cepeda:2019klc} and at the future $e^+e^-$ factories; in any case, the new experimental data, even if the search results are negative, will further exclude significant parts of these points.

Thus, the tentative 146 GeV Higgs boson signal seen by CMS can be interpreted within CP4 3HDM case $B_1$ and leads to experimentally testable predictions. Its characteristic leptonic decay signatures, with the $ee, \mu\mu, e\mu$ rates being of the same order of magnitude, offers a distinct experimental method that sets it apart from conventional Higgs scenarios, such as the 2HDM. Future collider BSM Higgs program should definitely include the $ee$ decay channel, an underappreciated probe at present.

\section{Summary}

In summary, the CP4 3HDM, a unique three-Higgs-doublet model with an exact CP symmetry of order 4, exhibits peculiar scalar and Yukawa sector properties that have been addressed in several studies. However, its impact on the lepton sector and lepton-scalar couplings has never been evaluated.
This is what we did in the present work. We compared our predictions and experimental constraints 
on the lepton flavor conserving and violating decays of the 125 GeV Higgs boson $\hsm$ 
and the LFV muon decays, both $\mu \to e\gamma$ and $\mu \to 3e$. Relying in the insights into the structure of the scalar and Yukawa sectors of the model
presented in the recent papers \cite{Liu:2024aew,Zhao:2025rds}, we set up an efficient scanning procedure and 
indeed found, in the scenario $B_1$, well-defined parameter space regions where the predictions are well within experimental limits, with a large margin.
In contrast, the other two CP4 scenarios, $B_2$ and $B_3$ typically lead to much larger $\mu \to e\gamma$ rates and are nearly excluded.

Now, after \cite{Zhao:2025rds} and the present work, there emerges a interesting conclusion:
the only CP4-invariant Yukawa sectors that are safely compatible with flavor and lepton observables
are $A$ or $B_1$ for the leptonic sector, $A$ for the down-quark sector, and $B_2$ for the up-quark sector.
It is at present unclear whether this observation is indicative of any structural feature of the model.

We conclude by stating that CP4 3HDM, with its peculiar symmetry and phenomenology, is still consistent with experimental data, 
at least in those scalar, quark, and lepton channels that have been considered so far.
Curiously, out of many CP4-invariant Yukawa sectors, only one is not excluded and is well within the limits.
Moreover, it leads to interesting predictions, such as the potentially large $\hnew \to ee$ signal.
However, we have not yet included the full body of experimental results available today.
Thus, it remains an intriguing question whether this unique version of the three-Higgs-doublet model
is indeed compatible with all searches and signals we have at present, or whether it will eventually be ruled out 
by any specific measurement not included in this work.
The fate of CP4 3HDM remains to be settled.

\section*{Acknowledgments}
This work was supported by Guangdong Natural Science Foundation (project No. 2024A1515012789).

\appendix

	\section{CP4 requires generation mixing}\label{CP4-defs}

A distinctive feature of higher-order $CP$ transformations is that they unavoidably mix
generations of the fields on which they act.
In the case of CP4 transformation $\phi_a \mapsto X_{ab} \phi_b^*$ acting in the scalar sector of the 3HDM,
we follow the suggestion of \cite{Ivanov:2015mwl,Ferreira:2017tvy} and choose the basis in which the matrix $X$ takes 
the following form:
\begin{equation}
	X =  \left(\begin{array}{ccc}
		1 & 0 & 0 \\
		0 & 0 & i  \\
		0 & -i & 0
	\end{array}\right)\,.
	\label{CP4-def}
\end{equation}
Note that applying this transformation twice leads to a non-trivial transformation in the space of doublets:
$\phi_{a} \mapsto (X X^*)_{ab} \phi_{b}$, with $XX^* = \mathrm{diag}(1, -1, -1)$. 
It is only when we apply CP4 four times that we get the identity transformation.
In contrast, any diagonal $X$ unavoidably generates a $CP$ symmetry of order 2, just because $XX^*$ gives the indentity.
This observation explain why no basis transformation can render $X$ of Eq.~\eqref{CP4-def} diagonal.

Next, in the Yukawa sector, we begin by introducing separate transformation matrices
for the left-handed lepton doublets and right-handed charged lepton singlets.
Then, one solves the consistency equations that guarantee that the leptonic Yukawa lagrangian in Eq.~\eqref{Yukawa-general}
remains invariant under CP4. These equations were solved in \cite{Ferreira:2017tvy}
and lead to four different cases labeled $A$, $B_1$, $B_2$, and $B_3$.
Case $A$ corresponds to the trivial scenario, in which all leptons couple only to the first Higgs doublet
and transform trivially under CP4; hence no LFV signals emerge. 
This possibility is always available within CP4 3HDM and was implicitly assumed in all previous works on CP4 3HDM. 
What we explore in this paper is whether any {\em non-trivial} CP4 scenario, $B_1$, $B_2$, or $B_3$ 
can be made compatible with the present LFV experimental constraints.

It must be stressed that the CP4 transformation acting on fermions also mix generations.
As a result, some elements of the Yukawa matrices $\Gamma_a$ are correlated. 
For example, in case $B_1$, the matrices are
\begin{equation}
	\Gamma_1 = \mmmatrix{0}{0}{0}{0}{0}{0}{g_{31}}{g_{31}^*}{g_{33}},\quad
	\Gamma_2 = \mmmatrix{g_{11}}{g_{12}}{g_{13}}{g_{21}}{g_{22}}{g_{23}}{0}{0}{0},\quad
	\Gamma_3 =  \mmmatrix{-g_{22}^*}{-g_{21}^*}{-g_{23}^*}{g_{12}^*}{g_{11}^*}{g_{13}^*}{0}{0}{0},
	\label{caseB1}
\end{equation}
where only $g_{33}$ needs to be real while all other parameters can be complex.
Expressions for the cases $B_2$ and $B_3$ can be found in \cite{Ferreira:2017tvy,Zhao:2023hws}.

\section{Yukawa coupling matrices $N_2$ and $N_3$} \label{expression-N2-N3}
For the sake of completeness, we outline here the expression for the Yukawa coupling matrices $N_2$ and $N_3$ 
for all three non-trivial CP4 leptonic sectors. These expressions coincide with the down-quark sector matrices derived in \cite{Zhao:2023hws},
with the obvious replacement of the masses and rotation matrices. 

In case $B_1$, the expression for $N_2^0$ in Eq.~\eqref{N0_23} can be written as
\begin{eqnarray}
	N_{2}^0 = M_e^0 \cot\beta - \frac{v}{\sqrt{2}s_\beta}\Gamma_1 = R_3^0 \cdot M_e^0 \,. \quad 
	R_3^0 = {\rm diag}(\cot\beta,\, \cot\beta, -\tan\beta)\,.
\end{eqnarray}
After the mass matrix diagonalization, we obtain $N_{2} = V_{L}^\dagger\, R_3^0 \, V_{L}\cdot M_e$.
As for $N_3^0$, we can present it as
\begin{equation}
	N_{3}^0 = \frac{1}{s_\beta} P_4 \cdot M_d^{0*} \cdot R_2 \,, \quad \text{where} \quad
	P_4 = \begin{pmatrix}
		0 & -1 & 0 \\
		1 & 0 & 0 \\
		0 & 0 & 0
	\end{pmatrix}, \quad
	R_2 = \begin{pmatrix}
		0 & 1 & 0 \\
		1 & 0 & 0 \\
		0 & 0 & 1
	\end{pmatrix}.
\end{equation} 
After the lepton field rotations, we get
\begin{eqnarray}
	N_{3} = \frac{1}{s_\beta} V_{L}^\dagger P_4 V_{L}^* \cdot M_e \cdot V_{R}^T R_2 V_{R}\,, 
	\label{N3-B1}
\end{eqnarray}
For case $B_2$, following the same logic, we write 
\begin{eqnarray}
	N_{2} &=& M_d \cdot V_{R}^\dagger\, R_3^0 \, V_{R}\,, \\
	N_{3} &=& \frac{1}{s_\beta} V_{L}^\dagger R_2 V_{L}^* \cdot M_e \cdot V_{R}^T P_4 V_{R}\,.
	\label{N23-B2}
\end{eqnarray}
Finally, for case $B_3$, the expressions are
\begin{eqnarray}
	N_{2} &=& M_d \cot\beta - \frac{1}{s_\beta c_\beta} \left( V_{L}^\dagger P_3 V_{L}^* \cdot M_e \cdot V_{R}^T P_3 V_{R} - V_{L}^\dagger P_4 V_{L}^* \cdot M_e \cdot V_{R}^T P_4 V_{R} \right)\,, \\
	N_{3} &=& \frac{1}{s_\beta} \left( V_{L}^\dagger P_4 V_{L}^* \cdot M_e \cdot V_{R}^T P_3 V_{R} + V_{L}^\dagger P_3 V_{L}^* \cdot M_e \cdot V_{R}^T P_4 V_{R} \right) \,.
	\label{N23-B3}
\end{eqnarray}
where $P_3 = {\rm diag}(0,0,1)$. 

\section{A general $U(3)$ rotation matrix} \label{appendix-U3}
We construct $U(3)$ rotation matrices $V_L$ and $V_R$ using the Chau-Keung form:
\begin{eqnarray}
	V &=& \begin{pmatrix}
		e^{i\psi_1} & 0 & 0 \\
		0 & e^{i\psi_2} & 0 \\
		0 & 0 & e^{i\psi_3}
	\end{pmatrix} 
	\begin{pmatrix}
		1 & 0 & 0 \\
		0 & c_{23} & s_{23} \\
		0 & -s_{23} & c_{23}
	\end{pmatrix}
	\begin{pmatrix}
		c_{13} & 0 & s_{13}e^{-i\delta} \\
		0 & 1 & 0 \\
		-s_{13}e^{i\delta} & 0 & c_{13}
	\end{pmatrix}
	\begin{pmatrix}
		c_{12} & s_{12} & 0 \\
		-s_{12} & c_{12} & 0 \\
		0 & 0 & 1
	\end{pmatrix} 
	\begin{pmatrix}
		e^{i\varphi_1} & 0 & 0 \\
		0 & e^{i\varphi_2} & 0 \\
		0 & 0 & e^{i\varphi_3}
	\end{pmatrix}.
	\label{SU3}
\end{eqnarray}
Here, we use shorthand notation $s_{ij}\equiv \sin\theta_{ij}, c_{ij}\equiv \cos\theta_{ij}$.
Since the simultaneous shift of all $\psi_i$ and all $\varphi_i$ does not change the result, we set $ \varphi_3 = 0 $.


\end{document}